\documentclass[final,5p,times,twocolumn]{elsarticle}

\usepackage[hidelinks]{hyperref}
\usepackage{graphicx}
\usepackage{amsmath}

\begin{document}

\begin{frontmatter}

\title{POEMMA-Balloon with Radio: a balloon-born multi-messenger multi-detector observatory }

\author[1]{Matteo Battisti\corref{cor1} }%
\ead{battisti@apc.in2p3.fr}

\author[2]{Johannes Eser}

\author[3]{Angela Olinto}

\author[4]{Giuseppe Osteria}

\author{for the JEM-EUSO Collaboration}

 \cortext[cor1]{Corresponding author}

\affiliation[1]{organization={ Université Paris Cité, CNRS, Astroparticule et Cosmologie, F-75013 Paris, France}} 
\affiliation[2]{organization={University of Chicago},  
                 city={Illinois}, 
                 country={U.S.A.}}
\affiliation[3]{organization={Columbia  University},  
                 city={New York}, 
                 country={U.S.A}}
                 
\affiliation[4]{organization={INFN Napoli}, 
                 country={Italy}}

\begin{abstract}
The Probe Of Extreme Multi-Messenger Astrophysics (POEMMA) is a proposed dual-satellite mission to observe Ultra-High-Energy Cosmic Rays (UHECRs), increasing the statistics at the highest energies, and Very-High-Energy Neutrinos (VHENs), following multi-messenger alerts of astrophysical transient events throughout the universe such as gamma-ray bursts and gravitational wave events. POEMMA-Balloon with Radio (PBR) is a scaled-down version of the POEMMA design, adapted to be flown as a payload on one of NASA’s sub-orbital Super Pressure Balloons (SPBs) circling over the Southern Ocean for up to 100 days after a launch from Wanaka, New Zealand. This overview will provide a summary of the mission with its science goals, the instruments, and the current status of PBR.
\end{abstract}

\end{frontmatter}


\section{The Science}
\label{sec:science}
The POEMMA-Balloon with Radio (PBR) is a scientific mission optimized for a flight on a NASA Super-Pressure Balloon (SPB), flying at a nominal altitude of 33~km.

PBR will explore multi-messenger science from a unique perspective (from sub-orbital altitudes), featuring a unique set of detectors and detection techniques, with a secondary goal to advance the technological readiness level (TRL) of a future multi-messenger space mission such as POEMMA \cite{POEMMA}.

\begin{figure}
    \centering
    \includegraphics[width=\columnwidth]{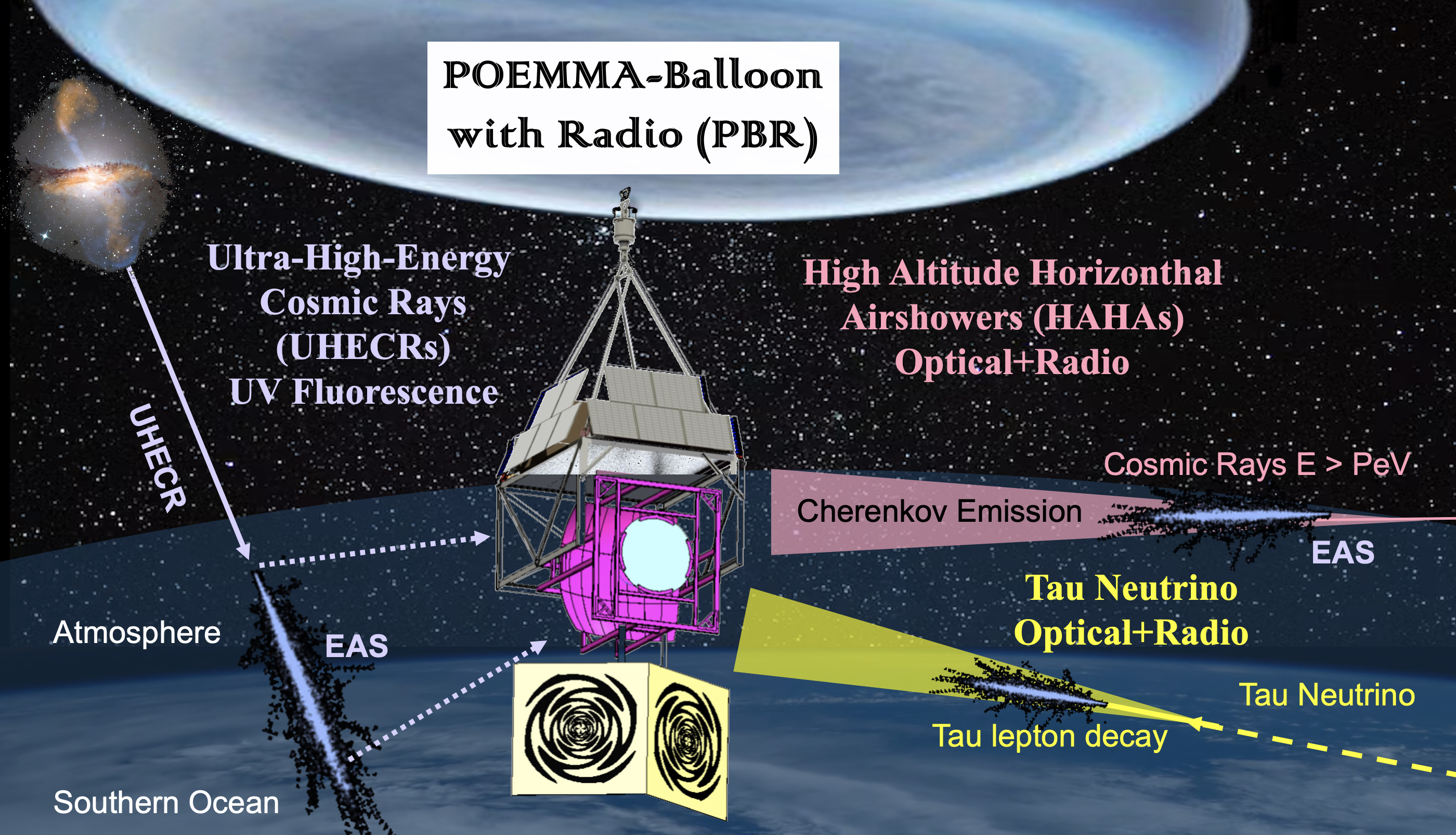}
    \caption{PBR design. The main types of researched events are highlighted, along with the main detector used to observe them.}
    \label{fig:PBR_design_and_science}
\end{figure}

The main science objectives of PBR are: (1) to observe UHECRs via the fluorescence technique from sub-orbital space; (2) to observe  High-Altitude Horizontal Air showers (HAHAs) with energies above the cosmic ray knee (E$>$0.5~PeV) using the optical and radio detection for the first time; and (3) to follow astrophysical event alerts in search of VHENs. The PBR design and main objectives are illustrated in Fig.~\ref{fig:PBR_design_and_science}.

\section{The detector}
\label{sec:detector}
PBR is built upon the experience of the previous balloon iterations, EUSO-SPB1 \cite{SPB1} and EUSO-SPB2 \cite{SPB2}.
The PBR instrument consists of a 1.1~m aperture Schmidt telescope
similar to the POEMMA design with two cameras in its focal surface: a Fluorescence Camera (FC) and a Cherenkov Camera (CC). In addition, PBR has a Radio Instrument (RI) optimized for the detection of EASs (covering the 50-500 Mhz range). The entire structure can freely rotate by 360$^\circ$ in azimuth and can be tilted from nadir up to 13$^\circ$ above the horizon in zenith.
The FC, devoted to the search of UHECR-induced EASs, is made of a matrix of 4 
Photo Detection Modules (PDMs) arranged in a 2$\times$2 configuration. A PDM consists of a 6$\times$6 array of 64-channel MAPMTs  
Photo Multi-Anode Photo-Multiplier Tubes (MAPMTs) [Hamamatsu R11265\footnote{For further details, \url{https://www.hamamatsu.com/content/dam/hamamatsu-photonics/sites/documents/99_SALES_LIBRARY/etd/R11265U_H11934_TPMH1336E.pdf}}, 64 pixels each] for a total of 9216 pixels imaged every 1.05~$\mu$s. The CC, devoted to the observation of cosmic-ray-induced HAHAs and search for neutrino-induced upward-going EAS, is made of a matrix of 32 Silicon Photo-Multiplier (SiPM) arrays [Hamamatsu S13361-3050NE-08\footnote{For further details, \url{https://www.hamamatsu.com/jp/en/product/optical-sensors/mppc/mppc_mppc-array/S13361-3050NE-08.html}}, 64 channels each] for a total of 2048 pixels. The CC covers a spectral range of 320-900~nm with an integration time of 10 ns.
The RI is mounted at the bottom of the telescope and moves solidly with the two cameras, resulting in a wide FoV that covers those of the FC and CC. It is based on the Low-Frequency (LF) instrument for the Payload
for Ultrahigh Energy Observations (PUEO) \cite{PUEO}.
In addition, a number of ancillary detectors will complete the PBR payload, including a set of infrared cameras for cloud monitoring and a charged particle, X-ray and gamma-ray detector that will work in conjunction with the CC. It will point in the same direction of the CC (overlapping FoV), with self-triggering capabilities as well as with the possibility to receive external triggers from the CC. It will be therefore possible to measure the shower emissions in the optical, radio, X and gamma bands at the same time.

\section{High-Altitude Horizontal Air showers (HAHAs)}
\label{sec:HAHAs}
\begin{figure}
    \centering
    \includegraphics[width=\columnwidth]{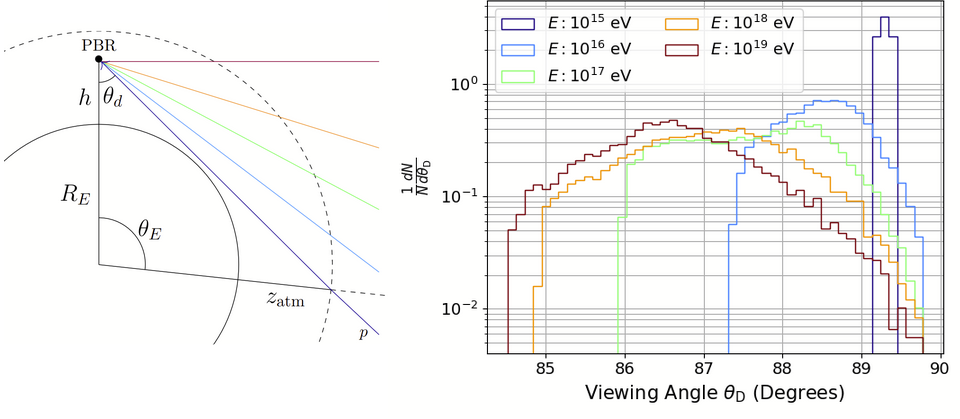}
    \caption{Left: HAHA trajectories with different inclination angles relative to PBR Nadir. Right: angular distribution of accepted HAHA events for different primary cosmic ray energies.}
    \label{fig:HAHAs}
\end{figure}
HAHAs refer to EASs induced by cosmic rays that skim the Earth’s atmosphere and traverse the telescope FoV, never intersecting the ground. PBR will be a unique laboratory to study the development of cosmic rays in such a peculiar condition. The majority of HAHA shower development occurs, in fact, above altitudes of 20~km, where the atmosphere is rarified, allowing for propagation over hundreds of kilometers (Fig.~\ref{fig:HAHAs}, left).
HAHAs will be observed by PBR in tilted mode, looking above the limb, mainly by the CC, with additional information provided by the RI, X and gamma-ray detectors.
PBR will observe HAHAs ranging from Earth’s limb (84.2°) to horizontal, at a rate of $\sim$1 event per minute. The Earth’s atmosphere acts as an energy filter, therefore the angular acceptance is energy dependent (Fig.~\ref{fig:HAHAs}, right). PBR might also be able to provide chemical identification on a statistical basis around the cosmic ray knee energy.

\section{Neutrino search from Targets of Opportunity}
\label{sec:ToO}
\begin{figure}
    \centering
    \includegraphics[width=.65\columnwidth]{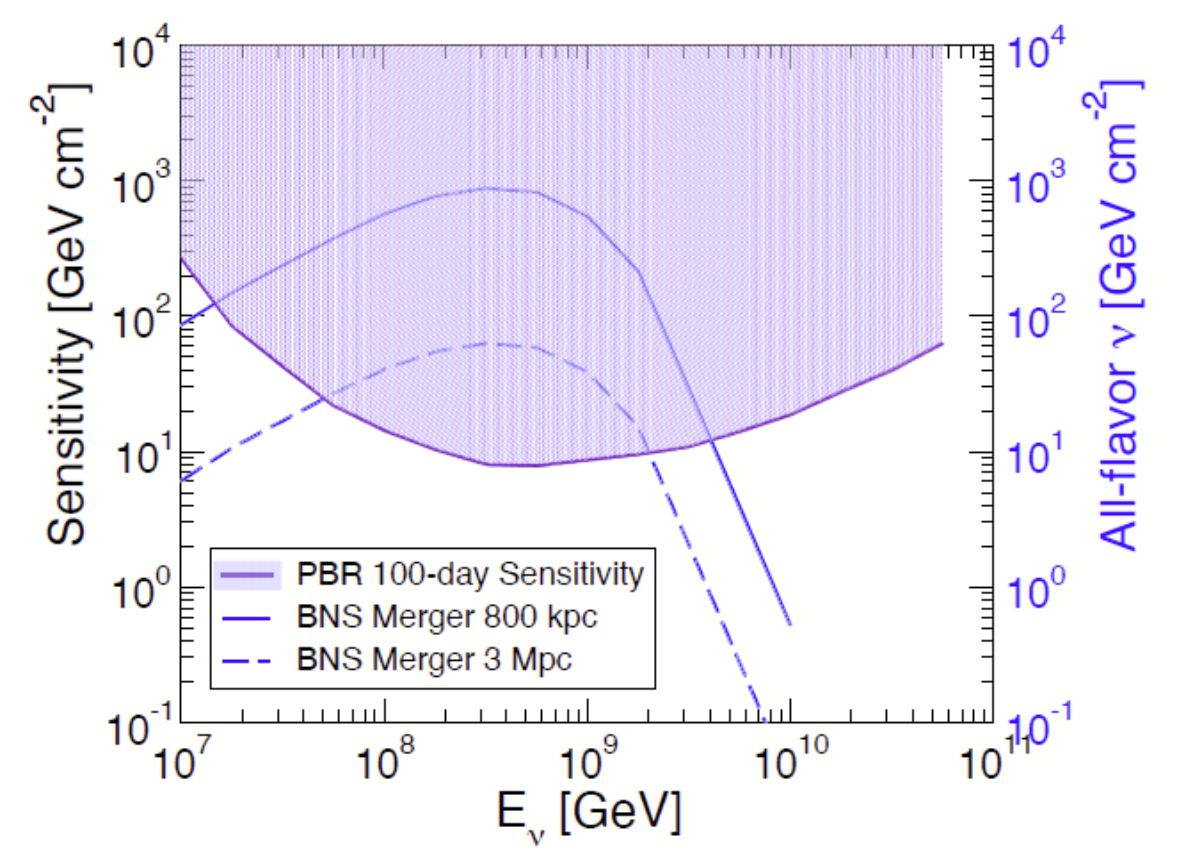}
    \caption{The potential all-flavor sensitivity range for PBR for a 100-day flight. 
    }
    \label{fig:ToO}
\end{figure}

With the Earth as a neutrino converter, high energy tau neutrinos can produce $\tau$-leptons that emerge from the Earth and initiate showers from their decay. When pointing below the limb, the CC can detect the Cherenkov light produced by the up-going showers \cite{ToO_for_SPB2} starting from $\sim$0.5~PeV (above 300~PeV the shower could be detected by the RI as well). Simulations show that the sensitivity of PBR to transient VHEN sources (supernovae, binary neutron star (BNS) mergers, tidal disruption events, blazar flares, and gamma-ray bursts) is comparable to current ground-based neutrino telescopes and can constrain some transient source scenarios (Fig.~\ref{fig:ToO}).

\section{Status and timeline}
PBR has been selected by NASA as part of the Astrophysics Research Analysis (APRA) Program.
The design of the instrument is well underway with the main elements already defined. The collaboration is currently going through the component's procurement, testing and prototyping phases. The full integration of the instrument is planned for early 2026, while the final field test will take place for a few months in the spring at the Telescope Array \cite{TA} site in Utah. The flight is planned for the spring of 2027 from NASA's Wanaka launch site in New Zealand, with an expected duration of up to 100 days.

\section*{Acknowledgements}

The authors would like to acknowledge the support by NASA award 80NSSC22K1488, by the French space agency CNES and the Italian Space agency ASI. We also acknowledge the invaluable contributions of the administrative and technical staffs at our home institutions.
This research used resources of the National Energy Research Scientific Computing Center (NERSC), a U.S. Department of Energy Office of Science User Facility operated under Contract No. DE-AC02-05CH11231.

  \bibliographystyle{elsarticle-num-names} 
  \bibliography{bibfileTemplate}

\begin{thebibliography}{6}
\expandafter\ifx\csname natexlab\endcsname\relax\def\natexlab#1{#1}\fi
\providecommand{\url}[1]{\texttt{#1}}
\providecommand{\href}[2]{#2}
\providecommand{\path}[1]{#1}
\providecommand{\DOIprefix}{doi:}
\providecommand{\ArXivprefix}{arXiv:}
\providecommand{\URLprefix}{URL: }
\providecommand{\Pubmedprefix}{pmid:}
\providecommand{\doi}[1]{\href{http://dx.doi.org/#1}{\path{#1}}}
\providecommand{\Pubmed}[1]{\href{pmid:#1}{\path{#1}}}
\providecommand{\bibinfo}[2]{#2}
\ifx\xfnm\relax \def\xfnm[#1]{\unskip,\space#1}\fi
\bibitem[{Olinto et~al.(2021)}]{POEMMA}
\bibinfo{author}{A.~V. Olinto}, et~al. (\bibinfo{collaboration}{POEMMA}),
\newblock \bibinfo{journal}{JCAP} \bibinfo{volume}{06} (\bibinfo{year}{2021}) \bibinfo{pages}{007}. \DOIprefix\doi{10.1088/1475-7516/2021/06/007}. \href{http://arxiv.org/abs/2012.07945}{{\tt arXiv:2012.07945}}.
\bibitem[{Abdellaou et~al.(2024)}]{SPB1}
\bibinfo{author}{G.~Abdellaou}, et~al.,
\newblock \bibinfo{journal}{Astroaarticle Physics} \bibinfo{volume}{154} (\bibinfo{year}{2024}) \bibinfo{pages}{102891}. \DOIprefix\doi{10.1016/j.astropartphys.2023.102891}.
\bibitem[{Eser et~al.(2023)Eser, Olinto, and Wiencke}]{SPB2}
\bibinfo{author}{J.~Eser}, \bibinfo{author}{A.~V. Olinto}, \bibinfo{author}{L.~Wiencke} (\bibinfo{collaboration}{JEM-EUSO}),
\newblock \bibinfo{journal}{PoS} \bibinfo{volume}{ICRC2023} (\bibinfo{year}{2023}) \bibinfo{pages}{397}. \DOIprefix\doi{10.22323/1.444.0397}. \href{http://arxiv.org/abs/2308.15693}{{\tt arXiv:2308.15693}}.
\bibitem[{Hughes et~al.(2023)}]{PUEO}
\bibinfo{author}{K.~A. Hughes}, et~al. (\bibinfo{collaboration}{"PUEO"}),
\newblock \bibinfo{journal}{PoS} \bibinfo{volume}{ICRC2023} (\bibinfo{year}{2023}) \bibinfo{pages}{1027}. \DOIprefix\doi{10.22323/1.44.1027}.
\bibitem[{Heibges et~al.(2023)}]{ToO_for_SPB2}
\bibinfo{author}{T.~Heibges}, et~al. (\bibinfo{collaboration}{JEM-EUSO}),
\newblock \bibinfo{journal}{PoS} \bibinfo{volume}{ICRC2023} (\bibinfo{year}{2023}) \bibinfo{pages}{1134}. \DOIprefix\doi{10.22323/1.444.1134}. \href{http://arxiv.org/abs/2310.12310}{{\tt arXiv:2310.12310}}.
\bibitem[{Tameda(2009)}]{TA}
\bibinfo{author}{Y.~Tameda},
\newblock \bibinfo{title}{{Telescope Array Experiment}},
\newblock \bibinfo{journal}{Nuclear Physics B - Proceedings Supplements} \bibinfo{volume}{196} (\bibinfo{year}{2009}) \bibinfo{pages}{74--79}. \DOIprefix\doi{https://doi.org/10.1016/j.nuclphysbps.2009.09.011}.

\end{thebibliography}

\end{document}